# Physics-Guided Exponential Model Design of High-Ge-Content SiGe Selective Epitaxy for Gate-All-Around Source/Drain Applications

Zhigang Li[1], Guobin Bai[1], Hengwei Cui[1], Wenlong Yao[1], Jianfeng Gao[1,a], Qifeng Jiang[1], Junjie Li[1], Junfeng Li[1], Yongliang Li[1], Huaxiang Yin[1], Xiaolei Wang[1], Jun luo[1]

[1]Institute of Microelectronics of The Chinese Academy of Sciences, Beijing, 100029, China

**Abstract:** High-germanium-content silicon-germanium (SiGe) epitaxy is critical for strain engineering in advanced gate-all-around (GAA) transistors. This paper demonstrates a physics-guided exponential function model that quantitatively links selective epitaxial growth (SEG) parameters to Ge incorporation kinetics in nanoscale trenches. By coupling surface diffusion limited transport, gradient strain, and competitive adsorption dynamics, the model predicts optimal conditions for bottom-up filling with maximal Ge content. For trenches with widths of approximately 60 nm, the optimized process achieved a maximum Ge content of 57.93% and demonstrated 100% selectivity against silicon nitride (SiN) and silicon dioxide (SiO$_2$). Cross-sectional TEM and EDS analyses reveal a graded Ge profile that minimizes interfacial defects and strain energy. Our results show that the established process-physics correlation will significantly facilitate the development of GAA devices with 5nm CMOS technology nodes and beyond.

---

[a]e-mail: gaojianfeng@ime.ac.cn (corresponding author)

# 1. Introduction

silicon germanium (SiGe) epitaxy growth has attracted attention in microelectronics, optoelectronics and thermoelectrics due to its low cost and feasibility[1-4]. Selective epitaxial growth (SEG) is one of the key front end-of-line (FEOL) process technologies today that has been used in complementary metal oxide semiconductor (CMOS) device manufacturing for 20 years[5]. Achieving a very high quality SEG SiGe is crucial for CMOS source and drain strain engineering where bandgap tailoring is essential for the mobility enhancement[6-8]. Selectivity and Ge content are important index to measure the quality of a deposited layer in SEG SiGe using the Reduced Pressure Chemical Vapor Deposition (RPCVD) process[9-12]. The Ge content of a thin film is related to many parameters, such as total flow rate, susceptor temperature, operating pressure and the Hydrogen chloride (HCl) partial pressure[13-15]. Although various analytical models relating Ge content to process parameters exist in the literature [16-18], their direct applicability to practical process optimization remains limited, and their physical interpretations are often unclear. Therefore, it is very important to address the problem of coupling multiple parameters and conducting optimization with physical insight. In this work, a physics-guided exponential function model is employed to analyze the complex coupling effects of different operating parameters on SEG SiGe. The process parameters derived from this model enable the Ge content of up to 57.93 at.%, which meets the requirements for source/drain selective epitaxy and strain engineering in advanced logic devices[6, 19, 20].

# 2. Experimental Section

The SEG SiGe layers were epitaxially grown on nano-trench pattern 8-inch wafers using RPCVD. Dichlorosilane ($SiH_2Cl_2$, DCS) and 10% germane ($GeH_4$) in $H_2$ served as Si and Ge precursors, respectively. HCl was utilized as the etchant to maintain selectivity during the epitaxy. Native oxide was removed via 100:1 HF wet clean. Prior to epitaxial growth, the nano-trench substrates underwent pre-cleaning via thermal treatment at a sufficiently high temperature (<850 °C) to prevent surface reconstruction of the microstructures.

Nano-trench Structure Fabrication: FEOL processes were utilized to fabricate the nano-trench with a SiGe/Si supper lattice. Firstly, a three cycles $Si_{0.7}Ge_{0.3}$/Si supperlattice film was epitaxially grown on the surface of the 8-inch silicon substrate, with a total film thickness controlled at approximately 100 nm. After the epitaxial growth, functional thin films such as plasma-enhanced silicon oxide (PEOX), amorphous silicon (a-Si), and $SiO_2$/SiN/$SiO_2$ (ONO) trilayer dielectric stack were sequentially deposited on the surface of the superlattice film. A photomask with 140 nm equidistant lines was used for photolithographic exposure and development. Subsequently, using the photoresist as a mask, the ONO and a-Si films were pattern-etched by a dry etching process, and finally, a linear pattern with a width of 220 nm was formed. A low-pressure silicon nitride (LPSiN) film with a thickness of 200 nm was grown by low-pressure chemical vapor deposition (LPCVD) technology, serving as the spacer of the device structure. Using the patterned film as a mask, deep etching was directly performed on the LPSiN spacer and the three cycles $Si_{0.7}Ge_{0.3}$/Si superlattice, and finally the target nano-trench structure was formed. A summary of process flow, a schematic feature and cross-section TEM of nano-trench with deposited SiGe on trench regions are shown Fig. 1(a) to 1(c). In-line cross-section Transmission Electron Microscopy (X-TEM) and Energy-Dispersive X-ray Spectroscopy (EDS) were typically used to characterize nano-trench layer properties.

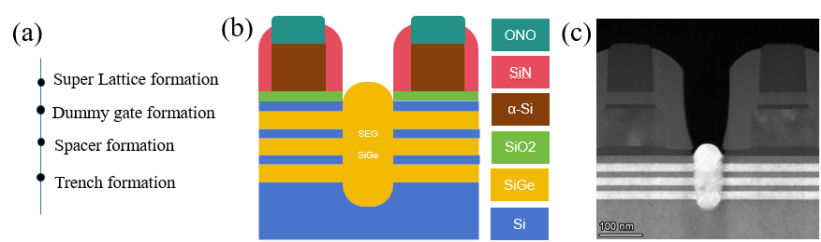

Fig. 1. (a) Process flow sequence, (b) a schematic view of SEG SiGe in nano-trench and (c) is a cross-sectional TEM image of a fully processed SEG SiGe nano-trench.

## 3. Results and discussion

**SEG Process Principle and Model:**

Selectivity of epitaxial layers is typically achieved by adding HCl to the growth chemistry, resulting in etching materials deposited on oxide and nitride surfaces. DCS is commonly used in SEG processes. DCS readily decomposes to form atomic silicon and gaseous HCl via a $SiCl_2$ intermediate adsorbed on a free Si-binding site. The schematic of SEG reaction, as shown in Fig. 2. This schematic illustrates a two-step surface reaction process for Ge incorporation onto a Si substrate, typical of atomic layer deposition or chemical vapor etching mechanisms:

(a) Dissociative Adsorption

In the dissociative adsorption stage, gaseous reactants including $GeH_4$, $H_2$, and HCl are introduced to the Si substrate surface. The molecule undergoes heterolytic bond cleavage, with bonds dissociating to release hydrogen radicals and deposit a Ge-containing intermediate onto active sites of the Si surface. Simultaneously, molecules interact with the substrate, enabling Cl atoms to adsorb onto Si surface sites, priming the substrate for subsequent reaction steps.

(b) Recombinative Desorption

Following surface adsorption, the adsorbed species undergo recombinative reactions to form volatile byproducts, which desorb from the surface. Specifically, surface-bound Ge intermediates react with adsorbed Cl atoms to form gaseous germylene dichloride (GeCl$_2$), while Si atoms on the substrate lattice combine with Cl to generate silicon dichloride (SiCl$_2$) vapor. Adsorbed hydrogen atoms also recombine to form molecules, and residual species are desorbed. This process results in the selective incorporation of Ge atoms into the Si surface, completing a cycle of surface modification.

The main reaction sequence is as follows[21, 22]:

$$\text{SiH}_2\text{Cl}_2 \text{ (g)} \leftrightarrow \text{SiCl}_2\text{(g)} + \text{H}_2 \text{ (g)} \tag{A1}$$

$$(*) + \text{SiCl}_2\text{(g)} \leftrightarrow \text{SiCl}_2(*) \tag{A2}$$

$$\text{SiCl}_2 (*) + \text{H}_2 \text{ (g)} \leftrightarrow \text{Si}(*) + 2\text{HCl (g)} \tag{A3}$$

$$\text{GeH}_4\text{(g)} + 2\text{HCl (g)} \leftrightarrow \text{GeCl}_2\text{(g)} + \text{H}_2 \text{ (g)} \tag{A4}$$

$$(*) + \text{GeCl}_2\text{(g)} \leftrightarrow \text{GeCl}_2(*) \tag{A5}$$

$$\text{GeCl}_2 (*) + \text{H}_2 \text{ (g)} \leftrightarrow \text{Ge}(*) + 2\text{HCl (g)} \tag{A6}$$

where (*) denotes a free silicon-binding site.

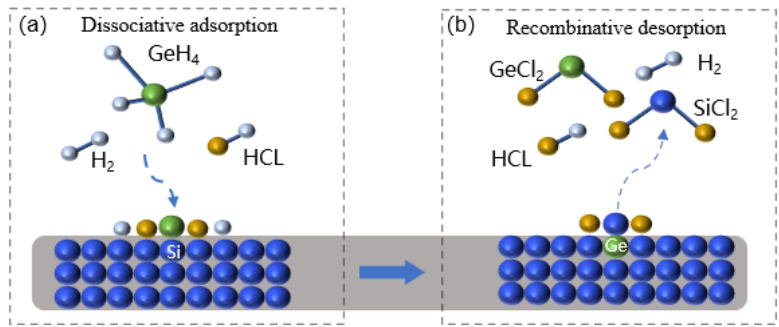

Fig. 2. Schematic of SEG reaction: (a) the dissociative adsorption, and (b) the recombinative desorption

One of the major quantities of interest in the deposition is the composition of Ge in the grown film and its dependence on growth conditions. For the SiH$_2$Cl$_2$/GeH$_4$ system for the intermediate temperature range (600 °C < T < 900 °C), a generally accepted relationship between the Ge content and the gas phase composition [23] [16] is

$$\frac{x^2}{1-x} = \text{cons} \tan t \times \frac{P_{GeH_4}}{P_{SiH_2Cl_2}} \tag{1}$$

Where P$_{GeH4}$ and P$_{SiH2Cl2}$ are the partial pressures of SiH$_2$Cl$_2$ and GeH$_4$, respectively.

But for SEG SiGe film grown form SiH$_2$Cl$_2$ /GeH$_4$ /HCl system, Equation (1) is not appropriate for process optimization. Since the reaction kinetics of the epitaxial process can be depicted by the Arrhenius equation, which takes the form of a natural exponential function, we thus adopted the fundamental exponential function model as the quantitative analysis model for the correlation between Ge composition and process parameters based on the physical principle.

We constructed an analytical model based on Arrhenius equation to explore the charateristics of SEG SiGe film, as follows,

$$y = y_0 - A_0 \cdot e^{-B_0 x} \tag{2}$$

Hence, it provides a helpful understanding of the SEG SiGe concentration and grow rate with respect to the process parameters. Consequently, the exponential function model parameters ($y_0$, $A_0$, $B_0$) have been extracted by fitting the raw data from the third part of the experimental data [14, 24, 25], as shown in Fig. 3(a) ~ (c). Using the drived fitting rules, seven process parameters were developed for SiGe selective epitaxy on blanket wafers. Subsequent X-ray diffraction (XRD) characterization confirmed that a Ge content of 50%, as shown in Fig. 3(d). The best-fitting model parameters ($y_0$, $A_0$, $B_0$) of the Model (2) are anticipated using the least-squares nonlinear curve regression analysis method. Model (2) was evaluated by $R^2$ as the coefficient of determination. This is a simple and efficient confidence level to evaluate the model fit. Using $R^2$ to evaluate the fitting emphasizes the accuracy of the prediction using the model curve. The minimum $R^2$ is greater than 0.99. The confidence level of each parameter is over 99%.

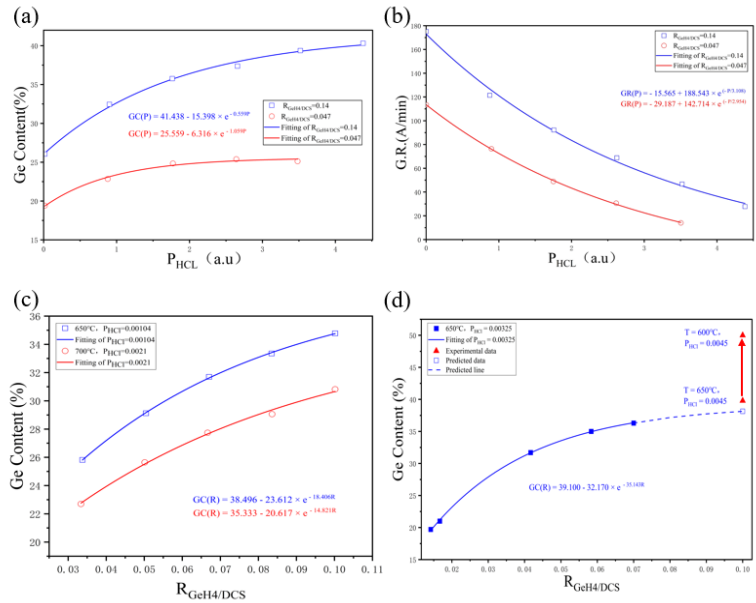

**Fig.** 3 Comparison between the raw data (scatter) and fitting in the Model (2) (solid line): (a) Fitting the Ge content with respect to HCl partial presure. (b) Fitting the Grow Rate with respect to HCl partial presure. The raw data in (a) (b) were adapted from the paper[25]. (c) Fitting the Ge content with respect to the ratio of GeH$_4$ to DCS. The raw data were adapted from the paper[14, 24]. (d) Fitting the Ge content with respect to the ratio of GeH$_4$ to DCS. The raw data were characterized by XRD.

As shown in Fig. 3, Ge content increases gradually with HCl partial pressure and approaching a saturation limit ($y_0$) for different GeH$_4$/DCS ratios whereas in the same time growth rate decreases. Saturation limit ($y_0$) has an important significance in process optimization. A decrease in the process temperature enables a substantial increase in the Ge content significantly.

To obtain the quantitative relationship between the SEG process parameters and Model (2) parameters, the mathematical relationship between the HCl partial pressure and $y_0$, $A_0$, $B_0$ was provided from expressions (3) – (6) in line with Origin. The mathematical relationship between the ratio of SiH$_4$ to DCS and $y_0$, $A_0$, $B_0$ was provided from expressions (7) – (9) . As indicated by expression (9), the Ge content approaches the saturation limit when $R$ reaches 0.1. Further adjustment of $R$ at this stage yields diminishing returns, whereas modulating the HCl partial pressure and process temperature offers an effective pathway to enhance the Ge content, as shown in Fig. 3(d).

$$GC(P) = 41.438 - 15.398 \times e^{-0.559P} \tag{3}$$

$$GC(P) = 25.559 - 6.316 \times e^{-1.059P} \tag{4}$$

$$GR(P) = -15.565 + 188.543 \times e^{-P/3.108} \tag{5}$$

$$GR(P) = -29.187 + 142.714 \times e^{-P/2.954} \tag{6}$$

$$GC(R) = 38.496 - 23.612 \times e^{-18.406R} \tag{7}$$

$$GC(R) = 35.333 - 20.617 \times e^{-14.821R} \tag{8}$$

$$GC(R) = 39.1 - 32.17 \times e^{-35.143R} \tag{9}$$

**Selectivity of the SEG Process：**

As reported in many literatures, selectivity is the primary challenge in the SEG processes, especially in GAA source/drain SEG[26-30]. Based on our experimental work and literature, selectivity issues can be categorized into four types, as illustrated schematically in Fig. 4.

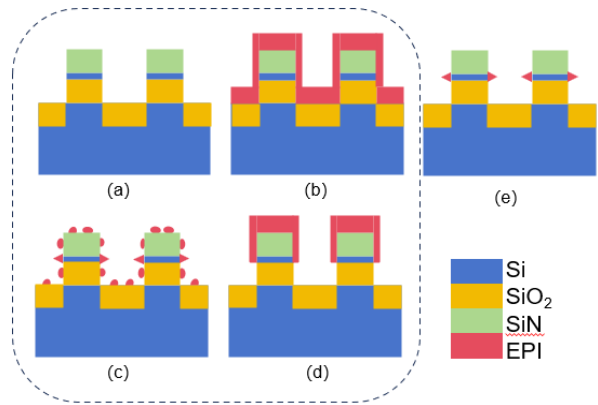

Fig. 4. Schematic of epitaxial selectivity issues: (a) over-selectivity, (b) non-selectivity, (c) under-selectivity, (d) biased-selectivity, (e) perfect-selectivity

The over-selectivity phenomenon arises from an excessively high partial pressure of HCl, which causes the nucleation rate to be lower than the etching rate and thus prevents any epitaxial film from growing on the surface, as shown in Fig. 4(a). The no-selectivity phenomenon is attributed to an excessively low partial pressure of HCl, which results in the etching rate being lower than the nucleation rate and thus allows a complete epitaxial film to grow even on the dielectric surface, as shown in Fig. 4(b). The under-selection phenomenon occurs when the partial pressure of HCl is close to the optimal ratio, which causes the etching rate to be slightly lower than the nucleation rate and thus leaves partial films remaining on the dielectric surface, as shown in Fig. 4(c). The biased-selectivity phenomenon is due to the fact that nucleation is more favorable on SiN, which results in the complete etching of the epitaxial film on $SiO_2$ while the film on SiN remains unetched, as shown in Fig. 4(d). Optimal HCl partial pressure achieves perfect-selectivity, with growth only on exposed silicon surfaces, as shown in Fig. 4 (e).

We identified the optimal partial pressure ratio of HCl via the model (2), resolved the aforementioned four selectivity-related issues, and achieved the optimal selectivity ratio, as shown in the Fig. 5. Crystalline SiGe materials are successfully grown on nano-trench while maintaining no growth on SiN and $SiO_2$ dielectrics. EDS mapping of the key elements (Si, Ge, O, C, N) provided the composition of the respective film layers in this nano-trench.

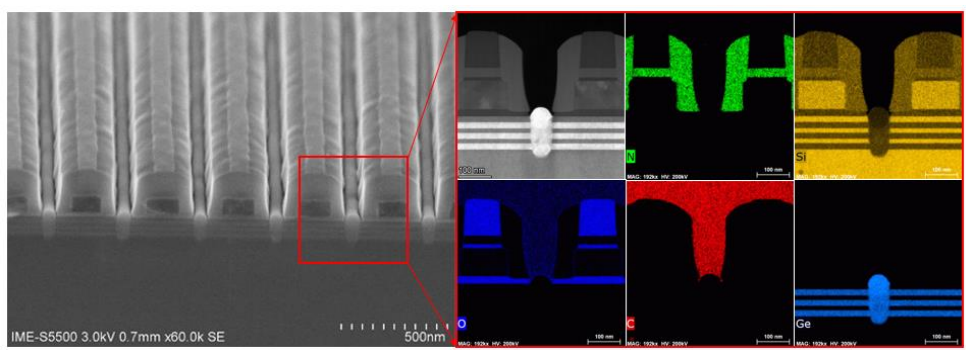

Fig. 5. Nano-trench with SEG SiGe, and EDS mapping of key elements

**Mechanism analysis of Nano-trench Filling in SiGe SEG:**

Fig. 6 shows X-TEM examples for the integration of a SiGe layer, which was processed in nano-trench. The SiGe growth happened from the different Si/SiGe sheets and from the Si bulk. Fig. 6(a) presents the cross-sectional TEM image of the nano-trench prior to SiGe SEG processing.

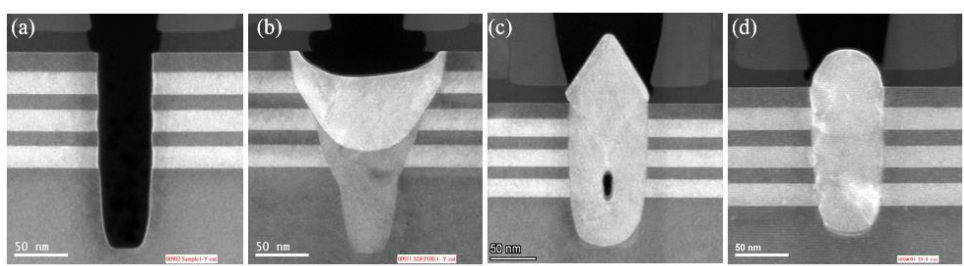

Fig. 6. X-TEMs of (a) a cross-sectional TEM image of the nano-trench before SiGe SEG, (b) the surface reconstruction of nano-trench after high $H_2$ bake temperature, (c) the void of SiGe

SEG in nano-trench, and (d) a cross-sectional TEM image of a fully processed SiGe SEG nano-trench.

The progression from (b) to (d) demonstrates how growth conditions can tune nanostructure morphology—from a wedge shaped structure, diamond shaped with an internal void, and finally to a rounded, strain-marked morphology—highlighting the sensitivity of nanostructure evolution to kinetic and thermodynamic growth parameters. The observed trench surface reconstruction in Fig. 6 (b) originates from thermal-driven surface diffusion and step-edge energy minimization under excessive H$_2$ bake temperatures (>850°C). The underlying physical mechanisms involve enhanced surface diffusion and surface energy minimization.

At elevated temperatures, the surface diffusion coefficient D$_s$ increases exponentially according to the Arrhenius relation[31]:

$$D_s = D_0 \exp(-\frac{E_a}{k_B T}) \tag{10}$$

When T > 850°C, $D_s$ becomes sufficiently large to enable mass transport of Si and Ge atoms from trench sidewalls to trench bottoms, as shown in the gray region at the trench bottom. When the bake temperature was below 850°C, the reconstruction phenomenon disappeared.

Under this circumstance, void formation was observed when the identical epitaxial process parameters as previously specified were employed, as shown in Fig. 6 (c). This void formation mechanism represents a classic "bread-loafing" effect in high-aspect-ratio trench filling, where kinetic limitations dominate over thermodynamic driving forces for conformal growth[32]. At low growth temperatures (<650°C), the surface diffusion length $\lambda_s$ becomes significantly shorter than the trench width $w$:

$$\lambda_s = \sqrt{D_S \cdot \tau_S} < w \tag{11}$$

where $\tau_s$ is the surface residence time.

The selective epitaxial growth of SiGe within nanoscale trenches involves complex interplay among surface diffusion, precursor adsorption/desorption kinetics, strain modulation, and interfacial energy minimization. The void phenomenon issue was resolved by implementing a processing scheme with a gradual increase in R$_{GeH4/DCS}$ and P$_{HCl}$ under the experimental conditions shown in Fig. 3(d). With increasing R$_{GeH4/DCS}$ and P$_{HCl}$ the etching effect of GeH$_4$ + HCl becomes more pronounced, while the growth rate decreases accordingly[13]. This provides sufficient time for surface atoms to diffuse toward the bottom of the trenches. Even if some atoms on the sidewalls do not reach the bottom through diffusion, they are still highly susceptible to removal by GeH$_4$ + HCl etching. Ultimately, the bottom-up filling of the nano-trench was successfully achieved, as shown in Fig. 6(d). The void-free, bottom-up filling achieved in ~ 60 nm narrow trenches demonstrates that this process is capable of realizing uniform epitaxy on the complex 3D surfaces of GAA suspended nanosheets/nanowires, which constitutes the foundation for fabricating low-resistance source/drain contacts.

**Characteristics of the SEG SiGe Film:**

For Semi-quantitative TEM-EDS line scan analyses were performed along the horizontal and vertical directions inside the nano-trench, and labeled "EDS1", "EDS2" and "EDS3", with the corresponding results presented in Fig. 7. Herein, the blue line denotes the Si atomic content, while the red line represents the Ge atomic content. EDS1 scan, taken near the top of the nano-trench, displays relatively stable Si and Ge contents across the nano-trench region, with minor fluctuations. The Si content remains almost as the same as Ge, about 50%. EDS2 scan, positioned at the mid-height of the nano-trench, reveals a distinct dip in both Si and Ge contents within the nano-trench's central region. The Ge content in the trench is about 40%. EDS3 scan shows a pronounced compositional gradient: Si content increases from the top to the bottom of the nano-trench, while Ge content increases from the top to the bottom. The gradually variation in $R_{GeH4/DCS}$ induces a graded strain field, which facilitates bottom-up filling of the nano-trench. Ge content in EDS3 within the nano-trench peaks at 57.93%, with an average value of 45.87%.

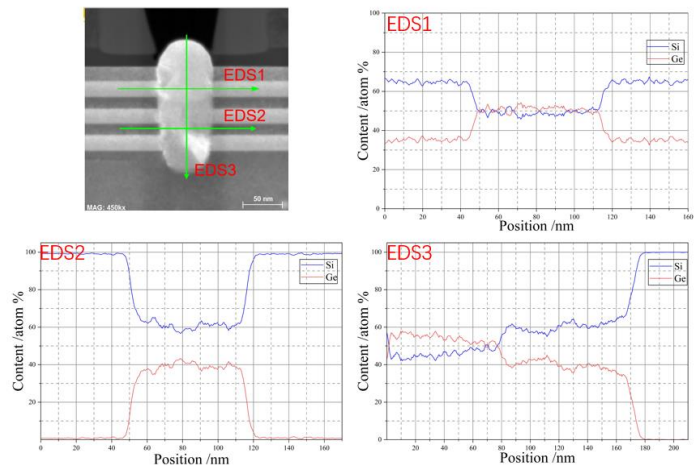

Fig. 7 EDS line scan analyses of SEG SiGe in the nano-trench

Fig. 8 presents the TEM and high-resolution TEM (HRTEM) images of the SEG SiGe interface in the nano-trench. A low magnification image of the TEM cross-section is shown in Fig. 8(a). The HRTEM image (Figs. 8(1, 2, 3)) from regions marked in (a) shows the good crystallinity of the epitaxy film near its interface. The inset in Fig. 8 (1, 2, 3) show the fast Fourier transform (FFT) of the image, which exhibits some streaks due to planar defects. A stack consisting of the SiGe epitaxial interface and the SiGe/Si superlattice film can be observed in the HRTEM image of Fig. 8(1, 2). A stack consisting of the SiGe epitaxial interface and the Si substrate can be observed in the HRTEM image of Fig. 8(3).

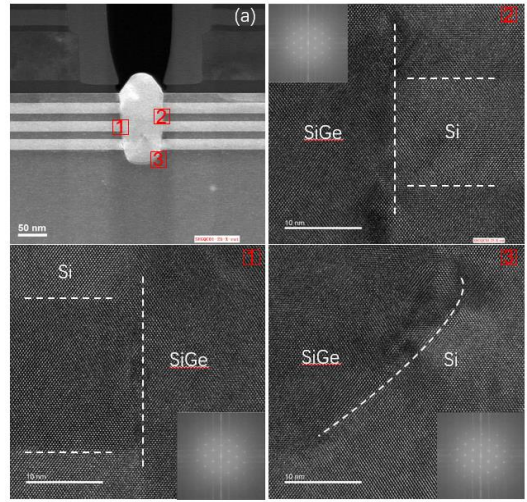

Fig. 8 Cross-sectional STEM image of nano-trench and the HRTEM images showing the interface of the SEG SiGe with the surface of the trench, respectively, including the SiGe/Si superlattice layer.

## 4. Conclusion

We have developed a physics-based exponential-function model that successfully guides the selective epitaxial growth of high-Ge-content SiGe in nanoscale trenches. The optimized SEG process achieved 100% selectivity against both SiN and $SiO_2$ dielectrics and delivered a maximum Ge concentration of 57.93%. Using the developed SiGe film, nano-trenches designed with $Si_{0.7}Ge_{0.3}$/Si superlattice were demonstrated in this study. The nanostructure evolution during the optimized epitaxy process were invesigated with HRTEM. The planar defects within the nano-trench were analyzed with FFT. Thus, a novel SEG process design model and SiGe film in nanoscale trench were reported, which could facilitate the development of source/drain selective epitaxy and strain engineering for GAA devices.

**Declaration of competing interest**

The authors declare that they have no known competing financial interests or personal relationships that could have appeared to influence the work reported in this paper.

**Acknowledgments**

This work was supported by the Beijing Natural Science Foundation (Grant No. Z220006).

**Data availability statements**

All data generated or analyzed during this study are included in this published article [and its supplementary information files].